\documentclass[aps,prl,twocolumn]{revtex4}

\usepackage{dcolumn}
\usepackage{graphicx,graphics}
\usepackage{amssymb}
\usepackage{rotate}
\usepackage{subfigure}
\usepackage{epsf}
\usepackage{epsfig}
\usepackage{amsmath,amssymb,amsfonts}
\usepackage{tabularx,hhline}
\usepackage{theorem}
\usepackage{enumerate,color}

\usepackage[normalem]{ulem}  

\newcommand{\eg}{\textit{e.g.} }

\begin{document}
\title{Analysis of a fully packed loop model arising in a magnetic Coulomb phase}

\author{L.~D.~C.~Jaubert, M. Haque, and R. Moessner}

\affiliation{Max-Planck-Institut f\"ur Physik komplexer Systeme, 
01187 Dresden, Germany.} 

\date{\today}

\begin{abstract}
The Coulomb phase of spin ice, and indeed the $I_c$ phase of water ice, naturally realise a fully-packed two-colour loop model in three dimensions. We present a detailed analysis of the statistics of these loops, which avoid themselves and other loops of the same colour, and contrast their behaviour to an analogous two-dimensional model. The properties of another extended degree of freedom are also addressed, flux lines of the emergent gauge field of the Coulomb phase, which appear as "Dirac strings" in spin ice. We mention implications of these results for related models, and experiments.
\end{abstract}

\pacs{75.10.Hk,75.10.Kt,75.60.Ch}

\maketitle

{\it Introduction:} In the study of magnetism, we are naturally led to consider local degrees of freedom and their correlations, such as the long-range order of the local spin orientation in a ferromagnet. In other settings, the fundamental degrees of freedom are extended, with polymers presenting perhaps the most familiar instance.

In magnets in two dimensions (2D), linearly extended objects do occur in the form of
magnetic domain walls. The study of such 'non-local' degrees of freedom,
involving questions such as: ``What is the probability for two bonds to be on
the same domain wall?'', is related to problems like the geometry of the
hull of a percolating cluster. Some  beautiful theories have been developed in
this context~\cite{Duplantier98a,Werner04a,Jacobsen98a}.
The same questions in higher dimension do not lend
themselves to similarly exact approaches, but fractal
extended objects in 3D have been studied in the contexts of polymer physics
\cite{Degennes79a}, cosmic strings~\cite{Vachaspati84a,Austin94a}, magnetic filaments in manganite materials~\cite{Viret04a}, and laser speckles \cite{OHolleran08a}.

Here, we discuss a 3D frustrated magnetic system, spin ice~\cite{Bramwell01a}, exhibiting
{\em two distinct} extended degrees of freedom, which we call loops and
worms. Spin ice is unusual in that its low-temperature magnetic state is
neither ordered (as in a ferromagnet) nor disordered (as in a paramagnet).
Rather, this state is a \emph{Coulomb phase}, where an emergent conservation
law leads to algebraic spin correlations at low temperatures \cite{Isakov04b}. This has recently been confirmed experimentally for the compound Ho$_{2}$Ti$_{2}$O$_{7}$ \cite{Fennell09a}.
The loops define a two-color fully packed loop model on the diamond lattice (the ``premedial'' lattice of the pyrochlore \cite{Henley10a}), and they also appear in models for pyrochlore compounds with two species of magnetic ions~\cite{Villain79a,Henley10a,Banks11a} or itinerant electrons subject to double exchange~\cite{Jaubert11a}.
Worms, named after their appearance in Monte Carlo worm algorithms~\cite{Barkema98a}, only come in one flavour and can pass through themselves and each other. They play a conceptually important role in the physics of spin ice as they are connected to ``Dirac strings'' and deconfinement in the Coulomb phase \cite{Castelnovo08a,Morris09a}.
Thus, besides their interest in the statistical mechanics of lattice models, we study loops and worms to elucidate the properties of this new magnetic phase.

\begin{figure}[t]
\centering\includegraphics[width=0.95\columnwidth]{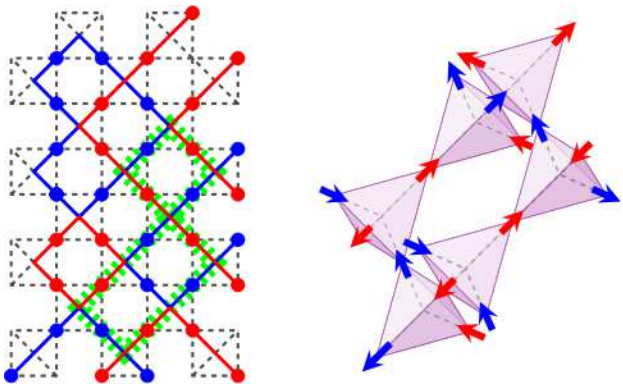}
\caption{ {\it Left:} Loops of two colors, blue (red) for up (down) spins, and
  a worm (dashed green, made of alternating up and down spins), on the
  checkerboard lattice.  The way we have drawn the loops highlights the fact
  that they fully occupy the premedial square lattice. 
A worm can cross
  itself and retrace its path. {\it Right:} Spin Ice model on the pyrochlore
  lattice with in (blue) and out (red) spins.  }
\label{fig:latt}
\end{figure}

In this paper, we numerically evaluate fundamental characteristics such as the probability distribution function (PDF) of loop length $\ell$, radius of gyration and the probability for two sites separated by distance $r$ to be on the same loop, $C(r)$. We present analytical arguments to account for the observed regimes and their concomitant power laws, and contrast this behaviour
to the analogous model in two dimensions, on the checkerboard
(Fig.~\ref{fig:latt}), whose premedial lattice is the square lattice.
Finally, we mention possible  experimental signatures and touch on
related models which naturally occur in the study of frustrated magnets and
multi-colored loop models \cite{Villain79a,Henley10a}.


{\em Loops and worms:} In spin ice, classical Ising spins
live on the sites of the pyrochlore, or equivalently, the
bonds of the 
diamond lattice.
At low temperature, an exponentially large number of degenerate ground state configurations is available, $\exp(N S_p)$, where $N$ is the number of spins and $S_p=\frac12\log\frac32$ is Pauling's ice entropy. These states obey the ice rules, stating that two of the bonds emanating from a diamond lattice site are occupied by up spins, the other two by down spins.
They correspond to the allowed configurations of cubic ice $I_c$, captured by the six-vertex model on the diamond lattice.
The ensemble of these states provides the Coulomb phase exhibiting an emergent gauge field and algebraic correlations~\cite{Isakov04b,Henley10a}.

By coloring the links occupied by up and
down spins differently, we obtain a fully packed (each link hosts a
loop segment) two-color loop model. 
In contrast, a worm contains an alternating sequence of adjacent up and down
spins (Fig.~\ref{fig:latt}). There are different possible constructions for a
worm.  We adopt an unbiased one in which a worm, having entered a tetrahedron,
exits through one of the two opposite spins in that tetrahedron with equal
probability.  A worm ends when it meets its initital site.

We consider periodic systems of square or cubic geometry with $L$ unit cells in each
direction, so that there are $N=4L^{2}$ and $N=16L^{3}$ lattice sites for the
checkerboard and pyrochlore cases respectively.  The smallest possible loop is
$\ell_{min}=4 \; (6)$ for checkerboard (pyrochlore).
We denote by $\langle \ell \rangle$ the average loop length and use subscripts $c$ or $2d$ ($p$ or $3d$) for checkerboard
(pyrochlore).

\begin{figure}[t]
\centering{\includegraphics[width=0.98\columnwidth]{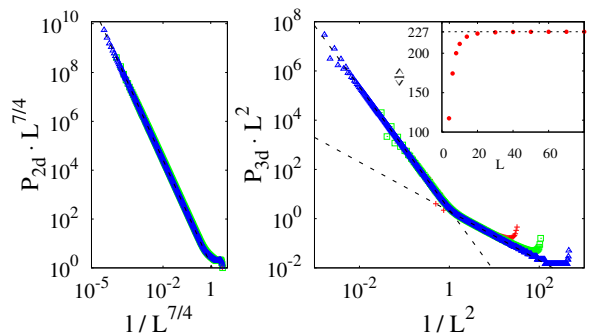}}
\caption{ Probability distribution functions of loop lengths $\ell$ on the
  checkerboard (left) and pyrochlore (right) lattice, for different system
  sizes ($L=$ 100, 400, 1000 in 2D and $L=$ 4, 14, 60 in 3D).
Each axis is scaled by a power of $L$.  
Note the absence of one-parameter scaling in 3D at large $\ell$.
 {\it Inset:} Average loop length on the pyrochlore lattice, converging
 with system size to a finite value of $227.3$ (dashed line).  }
\label{fig:pdf2d}
\end{figure}

\textit{Loop length distributions:}
Fig.\ \ref{fig:pdf2d} presents the PDF of the loop length obtained
using the Monte Carlo worm algorithm \cite{Barkema98a}.
For the checkerboard, the PDF has a single power-law behavior, $P_{2d}\sim
\ell^{-\tau_{c}}$, with $\tau_c = 2.14(1)$.  For related models (\emph{sans}
ice rules), this quantity is known exactly to be $\tau_c = 2 + 1/7$
(Ref.~\cite{Jacobsen98a} and Refs.\ thereof).  No such exact value is
available for $\tau_{p}$.  In fact, the situation for the 3D
pyrochlore is dramatically different.  There are now two different power-law
regions: a short loop region where the PDF scales as $\sim{L^3}\ell^{-2.50(1)}$,
and then a crossover at $\ell_{1}\sim{L^2}$ to a large-$\ell$ region with scaling
$\sim\ell^{-0.98(3)}$.
The second regime is due to the influence of \emph{winding loops}, which close
only after crossing the periodic boundaries.  We have checked
that the PDF of only non-winding loops has a single power law,
$\sim{L^3}\ell^{-5/2}$.  In 2D, the loop distribution is dominated by
non-winding loops at all $\ell$.  This is reminiscent of P\'olya's theorem,
that a random walk in 2D is \textit{recurrent} (always returns to the initial
point), while in 3D it is \textit{transient} (a finite probability never to return)~\cite{Hughes95a}.
In both 2D and 3D, $P(\ell)$ briefly increases just before the cutoff at large $\ell$.

\emph{Winding and non-winding loop fractions:} 
In 2D, the average number of winding loops is 1.86(1) for large
$L$.  In contrast, in the pyrochlore this number scales as
$N_{\rm w}\sim{\ln L}$.  In both cases the number of non-winding loop scales
as the total number of sites.
In the large-$L$ limit, the percentage of pyrochlore sites belonging to
non-winding and winding loops are $6.310(5)\%$ and $93.690(5)\%$ respectively.
Thus a finite but small portion of the system is covered by an extensive
number of short loops, while most of the lattice sites belong to a few large
loops.  

\begin{figure}[th]
\centering{\includegraphics[width=0.98\columnwidth]{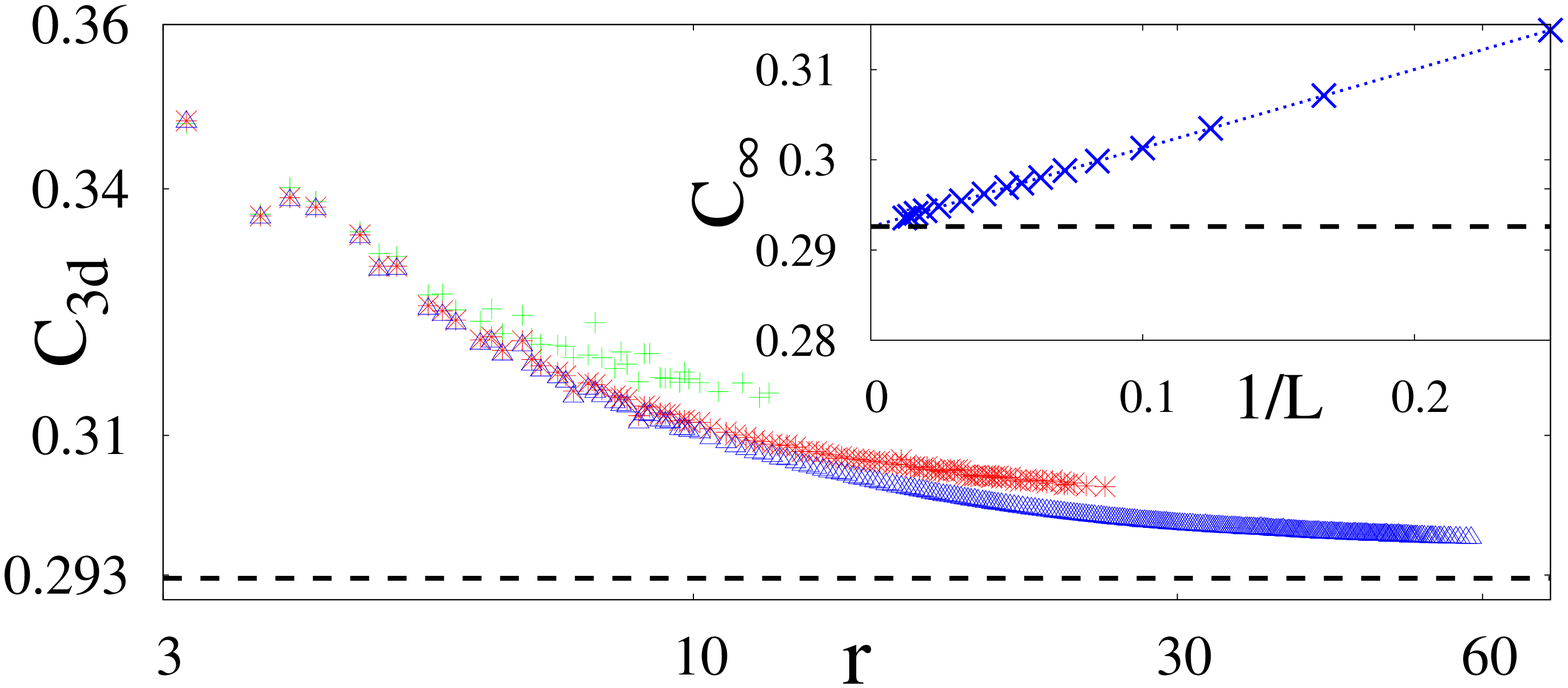}}
\caption{
Probability for two spins to be on the same loop as a function of the distance
between them, on pyrochlore lattices with $L=$ 4, 8, 18, asymptoting to $C_{\infty}=0.2926(10)$ (dashed lines), found by an extrapolation to large system sizes (inset).
}
\label{fig:2spin_proba_sameloop}
\end{figure}

\textit{Probability to be on the same loop:}
In 2D, we find $C(r){\sim}r^{-1/2}$, in analogy with~\cite{Jacobsen98a}.
By contrast, in 3D, the leading term in $C(r)$ is a constant
(Fig. \ref{fig:2spin_proba_sameloop}), i.e.\ sites far from each
other have a nonzero probability to be on the same loop.  This is another
qualitatively new feature of the 3D case arising from the transient nature of
3D loops.

\textit{Radius of gyration:} This is defined as
$R^{2}=\frac{1}{\ell}\sum_{i=1}^{\ell} \left|\mathbf{r}_{i}-\langle\mathbf{r}\rangle\right|^{2}$
where $\mathbf{r}_{i}$ is the position of the $i$th site of the loop of length
$\ell$ and $\langle\mathbf{r}\rangle=\sum_{i} \mathbf{r}_{i}/\ell$.  This follows a power law $R\sim \ell^{\nu}$. 
We find $\nu= 0.573(5)$ in 2D, consistent with 4/7 \cite{Jacobsen98a}. 
In 3D, we find $\nu=0.500(5)$.  This is the value for the random walk rather
than the self-avoiding ($\nu {\approx} 3/5$) walk, even though our loops are
self-avoiding.  This is similar in spirit to the observation for \emph{dense} polymer
solutions~\cite{Degennes79a}, that the need to avoid other loops counteracts the
effects of self-avoidance.
%

\textit{Ideal-chain analogy:}
We next present an analysis which accounts for the above results by applying
known results from the theory of random walks (ideal chains) to this setting.
The probability for an ideal chain starting at $\mathbf{x_{o}}$, to visit position $\mathbf{x}$ after $\ell$ steps, is
$p(\mathbf{x_{o}},\mathbf{x};\ell) = (2\pi\ell)^{-3/2}e^{-(\mathbf{x}-\mathbf{x_{o}})^2/2\ell}$, in the 3D continuum \cite{Hughes95a}.
Thus, the probability to come back to the initial point is $p(\mathbf{x_{o}};\ell) = \left(2\pi\ell\right)^{-3/2}$.
Summing over all possible starting positions, one obtains
the following loop length PDF for non-winding loops:

\begin{equation}
P_{3d}^{\rm nw}(\ell)\approx \dfrac{1}{\ell} \sum_{\mathbf{x}_{o}}\;p(\mathbf{x_{o}};\ell)\sim L^{3}\ell^{-5/2}
\label{eq:proba3dpdf}
\end{equation}
where the factor $1/\ell$ compensates the arbitrariness in defining the
starting position along the loop of size $\ell$.

To understand the different exponent for long loops, we have to consider
winding loops.  These may be regarded as loops that start at a point in the
``original'' sample but end at an equivalent point in a ``copy'' sample, arising from periodic boundary conditions. The copies cover all space.  The probability for the loop reaching the equivalent point in the $i$-th nearest-neighbor copy at distance $r_i=n_{i}L$, is $P(i;\ell)=(2\pi\ell)^{-3/2} e^{-r_i^2/2\ell}$ in 3D.
The number of copies which are $i$-th nearest neighbors scales as
$4\pi(r_{i}/L)^{2}$ for large $i$.  Thus the total probability for a winding
loop starting at the origin to be of size $\ell$ is
\begin{equation}
p(\mathbf{0};\ell)\approx 
\sum_{i=0}^{\infty}4\pi n_{i}^{2} 
\left(\dfrac{1}{2\pi \ell}\right)^{3/2}\exp\left[-\dfrac{n_{i}^{2}L^{2}}{2\ell}\right]
\approx \frac{1}{L^{3}}\,,
\label{eq:proba3dinf}
\end{equation}
after approximating the sum by an integral.  We note that $p(\mathbf{0};\ell)$ is independent of $\ell$ and scales as $L^{-3}$.
Summing over all possible initial points, the total probability to have a loop
of size $\ell$ becomes
\begin{equation}
P_{3d}^{\rm w}(\ell)\approx\dfrac{1}{\ell} \sum_{\mathbf{x_{o}}}\;p(\mathbf{0};\ell)\sim \ell^{-1}
\label{eq:proba3dlong}
\end{equation}
independent of system size $L$.
The crossover between the two behaviors occurs at the length scale
$\ell_{1}{\sim}L^2$, \emph{i.e.} when the radius of gyration of a loop reaches the
system size.

%
The average loop lengths $\langle\ell\rangle$ in both checkerboard and
pyrochlore cases are finite.  For the checkerboard, the PDF exponent $\tau_c$
being larger than 2 ensures a non-diverging $\langle\ell\rangle_c$.  With the
approximation that $P_{2d}(\ell)\propto\ell^{-\tau_c}$ at all even $\ell$
starting from $\ell_{min}=4$, one estimates
\begin{equation}
\langle \ell\rangle_{c}= \left(  \sum_{\ell=4,\ell\in
    2\mathbb{N}}^{+\infty}\;\ell.\ell^{-\tau_{c}} \middle/ \sum_{\ell=4,\ell\in
    2\mathbb{N}}^{+\infty}\;\ell^{-\tau_{c}}  \right)
~\approx~ 24.9  \,,
\label{eq:ellchess2}
\end{equation}
%
%
very close to the numerically obtained average 24.68(3).
%

For the pyrochlore, the combination of non-winding ($P_{3d}^{\rm
  nw}\sim{L^3}\ell^{-5/2}$) and
winding loops ($P_{3d}^{\rm w}\sim\ell^{-1}$) conspire to produce a remarkably large but finite average loop
length: $\langle\ell\rangle_p=227.5(5)$.
Even though they increase $\langle\ell\rangle_{p}$ by an order of magnitude
compared to the 2D model, winding loops do not manage to make it
divergent.

Considering non-winding and winding loops separately, we find that the
non-winding loop average saturates with $L$, to the value
$\langle\ell\rangle_{\rm n.w.} = 14.34(4)$.  (Eq.~(\ref{eq:ellchess2}) adapted
to the pyrochlore gives 15.3).  The winding loops by themselves have a
diverging average $\langle\ell\rangle_{\rm w} \sim L^3/\ln{L}$.

Concerning the
 probability for two sites to be on the same loop, 
we note that the number of spin pairs
belonging to the same loop of length $\ell$ is
$\tfrac{1}{2}\ell(\ell-1)\times P(\ell)$, while there is a total of $\tfrac{1}{2} N(N-1)$
pairs of spins in the system. The probability averaged over all pairs is thus
\begin{equation}
\overline{C(r)}  \sim  \int_{6}^{8\,L^{3}}\dfrac{\ell(\ell-1)}{N(N-1)}\,P_{3d}(\ell){\rm d}\ell \sim \mathcal{O}(L^{0})
\end{equation}
The non-winding loop contribution to this integral vanishes at large $L$, but
the winding loop contribution leads to a constant term
$C_{\infty}=C(r\rightarrow \infty)$ (Fig.~\ref{fig:2spin_proba_sameloop}).
Indeed, we note that there
will be several loops in a large system, each of which contains a finite
fraction, $f_{i}$, of all sites.  (We found $f_{0}\approx 0.41, f_{1} \approx
0.30, f_{2} \approx 0.12$ \dots).

The 'scaling function' displayed in Fig.~\ref{fig:pdf2d} shows that in 3D, there are in fact two length scales, namely when the radius of gyration hits the system size, $l_1\sim L^2$, and when the loop explores the full volume, $l_2\sim L^3$. The behaviour for $l>l_1$ obviously is influenced by the nature of the boundary conditions, e.g.\ the loops winding many times can break up into several loops terminating on the surface when open boundaries are considered.

For 2D, we note an intriguing similarity to the geometry of percolation hulls, which have a size distribution scaling with the same exponent we find above~\cite{Saleur87a}. This we can rationalise as follows. Our loops can be thought of as defining a specific percolation problem, with checkerboard lattice sites painted in either colour with probability $p=1/2$. It has been argued that neither short-range nor sufficiently rapidly decaying algebraic correlations between the occupied bonds  influence the percolation critical exponents \cite{Weinrib83a}. The 2D spin ice correlations indeed decay sufficiently fast, as $r^{-2}$. An additional feature of our loop ensemble is that loops of the same colour cannot cross or branch. From this it follows that each loop is at the same time a hull, whose length/perimeter should scale as $L^{7/4}$ (Fig.~\ref{fig:pdf2d}).

%

\begin{figure}[t]
\centering\includegraphics[width=8cm]{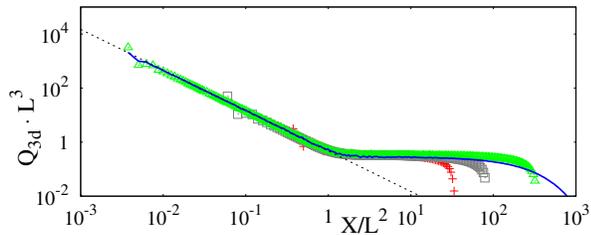}
\caption{ 
Data points: probability distribution of number of spins flipped by worms
($L=$ 4, 10, 40).  
Solid line: probability distribution of full worm length ($L=40$) which weighs each site by the number of times it has been visited. This departs parametrically only for the longest worms, where multiple visits to a site become appreciable.
}
\label{fig:5}
\end{figure}

\textit{Worms:}
We now address properties of the worms. These are efficiently evaluated as our
Monte-Carlo algorithm is in fact based on constructing worms, as flipping all
spins in a worm conserves the ice-rules.  Because worms can retrace parts of
their path, their actual length can be longer than the system size.  We
therefore look at the distribution $Q(X)$ of the number of spins flipped
during a worm update (Fig.~\ref{fig:5}), i.e., the number of sites visited by the worm an odd number of times.  Again, two regimes appear in 3D, an $\ell^{-3/2}$ behavior (numerical exponent -1.48(4)) for $\ell{\lesssim}\ell_{1}$, and a $\ell$-independent region for
larger $\ell$.  These ($-3/2$ and 0) are random-walk exponents that can be
derived similarly as in the loop case, Eqs.\ \eqref{eq:proba3dpdf} and
\eqref{eq:proba3dlong}.  The exponents for $Q(X)$ are shifted from the loop $P(\ell)$
exponents ($-5/2$ and $-1$) by 1; indeed if one asked the question ``How long is the loop on which a randomly chosen spin sits ?'', one would also get exponents $-3/2$ and 0.

The winding worm regime implies that 3D worms on average visit a finite
fraction of the system size ($\approx$ 15\%). Hence, a worm algorithm where a Monte Carlo
update is made by flipping the spins in a worm only requires a finite (typically $5 - 10$)
{\em constant} number of worms to decorrelate the system in 3D, \textit{independent} of system size $L$, while the different behaviour of winding worms in 2D requires an
increasing number of worm updates for decorrelation with $L$.

\textit{Conceptual and experimental implications:} 
For both 2D and 3D, we have recovered exponents originally derived for systems
\emph{without} ice rules.  The divergence-free conditions of Coulomb phases seem to
have little effect on the 'universal' loop and worm statistics we have considered, which was
not \emph{a priori} to be expected, as the ice-rules impose algebraic spin correlations.

In the context of spin ice, worm statistics plays an important conceptual role
as worms describe the tension-free (emergent) flux loops which are
characteristic of the Coulomb phase. Here we have shown that their statistics resembles that of an ideal chain. Since, in a magnetic field, the worms have been utilised as an experimental
diagnostic of Coulomb phase physics~\cite{Morris09a}, a more detailed study
of their statistics as a function of field strength would be worthwhile.

Regarding the loops, these are most naturally probed as a non-local
correlation function. For instance, for a system of electrons which can hop
only along a loop, the existence of the winding loops shows up
in conductivity properties, as we will discuss
elsewhere~\cite{Jaubert11a}.
%

To the best of our knowledge, our loops first appeared in Villain's seminal paper in the context of a model of pyrochlore Heisenberg magnets with two species of magnetic ions~\cite{Villain79a,Henley10a,Banks11a}. Villain already noted the possibility of two (recurrent and transient) loop populations, which also show up for cosmic strings~\cite{Vachaspati84a,Austin94a} and laser speckles~\cite{OHolleran08a}. 

In Villain's model, each loop has a distinct colour encoding the direction of its Heisenberg spin (which is continuously variable and hence the number of colours is infinite), and spins are
correlated only if they belong to the same loop. A finite value of
$C_\infty$ thus implies long-range spin order.  
Here we comment on two noteworthy features. Firstly the presence of several loops of
size $\mathcal{O}(L^{3})$ implies coexistence of several spatially
inter-twined but independent populations of long-range ordered
spins. Secondly, a short-range interacting classical spin Hamiltonian of the kind envisaged by Villain allows gapless excitations for the Heisenberg spins. Thermal
fluctuations out of this set of states could lead to
locking of these populations into a more conventional collinear ordered
structure.  This is clearly a productive field for more detailed
studies.

If one assigns a weight to each loop reflecting the number of flavours, $n$, it can take (which is not what is done in Villain's model, whose colours are used to distinguish the loops, not to weigh them), it becomes more favourable to have many, short loops as $n$ grows.
There is an extended literature on loop models, with some 3D work~\cite{Shtengel05a, Ortuno09a}, from which it is known that  a phase transition results.
In our case, an $n=\infty$ state is the ``hexagonal protectorate'' loop crystal proposed in the context of magnetodistortive phase transitions in the spinel compound ZnCr$_{2}$O$_{4}$~\cite{Lee02a}. This has {\em only} loops of length 6 and breaks lattice translation symmetry.

In summary, we have presented an analysis of a set of extended degrees of freedom arising in an exotic phase of a three-dimensional magnet. This we hope will be of dual interest both from a statistical mechanics and a magnetic materials relaxation perspective. Further studies of these models, \eg a quantum version thereof, are ongoing~\cite{Shannon11}.

\textbf{Acknowledgments $-$} We thank John Chalker, Chris Henley and Vincent Pasquier for useful discussions.


\begin{thebibliography}{10}

\bibitem{Duplantier98a}
B.~Duplantier, Phys.\ Rev.\ Lett.\ {\bf 81},  5489  (1998).

\bibitem{Werner04a}
W.~Werner,  in {\em Random planar curves and Schramm-Loewner evolutions}  (Springer, 2004), Chap.~2.

\bibitem{Jacobsen98a}
J.~L.~Jacobsen and J.~Kondev, Nucl.\ Phys.\ B {\bf 532},  635  (1998).

\bibitem{Degennes79a}
P.~G.~de~Gennes, {\em Scaling concepts in polymer physics} (Cornell Univ.  Press, Ithaca, 1979).

\bibitem{Vachaspati84a}
T.~Vachaspati and A.~Vilenkin, Phys.\ Rev,\ D {\bf 30},  2036  (1984).

\bibitem{Austin94a}
D.~Austin, E.~J.~Copeland, and R.~J.~Rivers, Phys.\ Rev,\ D {\bf 49},  4089  (1994).

\bibitem{Viret04a}
M.~Viret {\it et~al.},  Phys.\ Rev.\ Lett.\ {\bf 93}, 217402   (2004).

\bibitem{OHolleran08a}
K.~O'Holleran, M.~R.~Dennis, F.~Flossmann and M.~J.~Padgett, Phys.\ Rev.\ Lett.\ {\bf 100}, 053902 (2008). 

\bibitem{Bramwell01a}
S.~T.~Bramwell, and M.~J.~P.~Gingras, Science {\bf 294}, 1495 (2001).

\bibitem{Isakov04b}
S.~V.~Isakov, K.~Gregor, R.~Moessner, and S.~L.~Sondhi, Phys.\ Rev.\ Lett.\  {\bf 93},  167204  (2004).

\bibitem{Fennell09a}
T.~Fennell {\it et~al.}, Science {\bf 326},  415  (2009).

  
\bibitem{Henley10a}
C.~L.~Henley, Annual Review of Condensed Matter Physics {\bf 1},  179  (2010).

\bibitem{Villain79a}
J.~Villain, Zeitschrift F\"ur Physik B {\bf 33},  31  (1979).

\bibitem{Banks11a}
S.~T.~Banks, S.~T.~Bramwell, T.~Fennell, M.~J.~Harris, unpublished.

\bibitem{Jaubert11a}
L.~D.~C.~Jaubert, M.~Haque, S.~Pitaecki, and R.~Moessner, in preparation  (2011).

\bibitem{Barkema98a}
G.~T.~Barkema and M.~E.~J.~Newman, Phys.\ Rev.\ E {\bf 57},  1155  (1998).

\bibitem{Castelnovo08a}
C.~Castelnovo, R.~Moessner, and S.~L.~Sondhi, Nature {\bf 451},  42  (2008).

\bibitem{Morris09a}
D.~J.~P.~Morris {\it et~al.}, Science {\bf 326},  411  (2009).

\bibitem{Hughes95a}
B.~D.~Hughes, {\em Random walks and random environments} (Oxford University Press,  1995).

\bibitem{Saleur87a}
 H.~Saleur, and B.~Duplantier, Phys. Rev. Lett. {\bf 58}, 2325 (1987). 
  
\bibitem{Weinrib83a}
A.~Weinrib, and B.~I.~Halperin, Phys. Rev. B {\bf 27}, 413 (1983). 

\bibitem{Shtengel05a}
K.~Shtengel and L.~P.~Chayes, J.~Stat.\ Mech.\  P07006  (2005) and references therein.

\bibitem{Ortuno09a}
M.~Ortu\~no, A.~M.~Somoza, and J.~T.~Chalker, Phys.\ Rev.\ Lett.\ {\bf 102},
070603 (2009). 

\bibitem{Lee02a}
S.~H.~Lee {\it et~al.}, Nature {\bf 418},  856  (2002).

\bibitem{Shannon11}
N.~Shannon, private communication

\end{thebibliography}
\end{document}